 \documentstyle[epsf,12pt]{article}
 
\newcommand{\be}{\begin{equation}}
\newcommand{\ee}{\end{equation}}
\newcommand{\bea}{\begin{eqnarray}}
\newcommand{\eea}{\end{eqnarray}}
\newcommand{\beann}{\begin{eqnarray*}}
\newcommand{\eeann}{\end{eqnarray*}}
\newcommand{\beasn}{\begin{sneqnarray}}
\newcommand{\eeasn}{\end{sneqnarray}}

\newcommand{\al}{\alpha}
\newcommand{\e}{\epsilon}

\def\bfsigma{\mbox{\boldmath $\sigma$}}

\catcode`@=11
\def\dif{{\rm d}}
\def\deriv{\@ifnextchar[{\@deriv}{\@deriv[]}}
   \def\@deriv[#1]#2#3{\mathchoice%
{{\dif^{#1}#2\over\dif{#3}^{#1}}}{{\dif^{#1}#2/\dif{#3}^{#1}}}%
{{\dif^{#1}#2\over\dif{#3}^{#1}}}{{\dif^{#1}#2/\dif{#3}^{#1}}}}

\def\presup#1{{}^{#1}\kern-.15em\relax}      
\def\presub#1{{}_{#1}\kern-.12em\relax}      

\def\secteqno{\@addtoreset{equation}{section}%
\def\theequation{\thesection.\arabic{equation}}}
\def\endsecteqno{\def\theequation{\@ifundefined{chapter}%
{\arabic{equation}}{\thechapter.\arabic{equation}}}}
\newcounter{subequation}
\def\thesubequation{\alph{subequation}}
\def\sneqnarray{\stepcounter{equation}\let\@currentlabel=\theequation
\setcounter{subequation}{1}
\def\@eqnnum{{\rm (\theequation\thesubequation)}}
\global\@eqcnt\z@\tabskip\@centering\let\\=\@eqncr\let\@@eqncr=\@@sneqncr
$$\halign to \displaywidth\bgroup\@eqnsel\hskip\@centering
 $\displaystyle\tabskip\z@{##}$&\global\@eqcnt\@ne
 \hskip 2\arraycolsep \hfil${##}$\hfil
 &\global\@eqcnt\tw@ \hskip 2\arraycolsep $\displaystyle\tabskip\z@{##}$\hfil
  \tabskip\@centering&\llap{##}\tabskip\z@\cr}
\def\endsneqnarray{\@@sneqncr\egroup $$\global\@ignoretrue}
\def\@@sneqncr{\let\@tempa\relax
   \ifcase\@eqcnt \def\@tempa{& & &}\or \def\@tempa{& &}
   \else \def\@tempa{&}\fi
     \@tempa \if@eqnsw\@eqnnum\stepcounter{subequation}\fi
     \global\@eqnswtrue\global\@eqcnt\z@\cr}
\def\nobiblabels{\def\@lbibitem[##1]##2{\@bibitem{##2}}}
\catcode`@=12

\secteqno
\begin{document}

\def\s{\sigma}
\def\g{\gamma}
\def\m{\mu}
\def\n{\nu}
\def\d{\delta}
\def\als{\alpha_{s}}


\title{{\bf Matching at one loop for the
      \\ four-quark operators in NRQCD}}

\author{{\sc A.\,Pineda} \, and \,
        {\sc J.\,Soto}\\
        \small{\it{Departament d'Estructura i Constituents
               de la Mat{\`e}ria}}\\
        \small{\it{and Institut de F{\'\i}sica d'Altes Energies.}}\\
        \small{\it{Universitat de Barcelona, Diagonal, 647}}\\
        \small{\it{E-08028 Barcelona, Catalonia, Spain.}}\\
        {\it e-mails:} \small{pineda@ecm.ub.es, soto@ecm.ub.es} }

\date{\today}

\maketitle

\thispagestyle{empty}

\begin{abstract}

The matching coefficients for the four-quark operators in NRQCD (NRQED) are
calculated at one loop using dimensional regularization for ultraviolet and infrared
divergences.
The matching for the electromagnetic current follows easily from our results.
Both the unequal and
equal mass cases are considered.
The role played by the Coulomb infrared singularities is explained in detail.

\end{abstract}

\medskip

Keywords: Effective Field Theories, NRQCD, NRQED, HQET, Matching.

PACS: 12.38.Bx, 12.20.Ds, 14.40.Gx.

\vfill
\vbox{
\hfill February 1998\null\par
\hfill UB-ECM-PF 97/16}\null\par

\clearpage



\section{Introduction}
\indent

\bigskip

 Effective field theories (EFTs) have become increasingly popular in
describing processes where
several scales are involved. In
particular, two EFTs, namely Heavy Quark Effective Theory (HQET)
and Non-Relativistic QCD (NRQCD) have been used for systems with heavy quarks. These
EFTs take advantage of the fact that the masses of the heavy quarks (charm and
bottom) are much
larger than the remaining dynamical scales in the problem.

\medskip

HQET was designed to study systems
with one
heavy quark \cite{hqet,R1,Neubert} and has become a standard tool
during the last years. Apart from the mass of the heavy quark ($m$) the remaining
dynamical scales in heavy-light systems reduce to a single one $\Lambda_{QCD}$.
 The HQET Lagrangian can
be organized in a power series of the inverse pole mass of the heavy quark. Each term in
this series consists of a gauge invariant operator.  Only two kinds of terms turn out to be
  important for heavy-light systems: (i)  terms containing light degrees of freedom
(gluons and light quarks) only
 (which are irrelevant in most of the phenomenological applications), and
(ii) terms containing
a bilinear in the
heavy quark fields.
 The size of each term is easily
estimated by assigning the scale $\Lambda_{QCD}$ to whatever is not a heavy mass in the
Lagrangian.

\medskip

NRQCD was designed to study systems with a heavy quark and a heavy antiquark
\cite{Lepage,Lepage2} and, although it is older than HQET, it has not received much
attention until recently \cite{Manohar}-\cite{Labelle}. In this case,
apart from the heavy quark mass, there are at least
two dynamical scales. Namely the typical relative momentum in the bound state ${\bf p}$
and the typical binding energy $E$.
Because of the existence of these two scales, the power counting rules
are different from the HQET case and the size of each term in the NRQCD Lagrangian
is not unique. Nevertheless, counting rules have been given to
estimate the leading size of each term (see
\cite{Lepage2}). Independently of the relative size of each term,
the NRQCD Lagrangian also consists of a power series
of the inverse pole mass of the heavy quark. Here, though, there are important terms of three
 kinds: the two first kinds correspond exactly to (i) and (ii) in HQET, where we include in
(ii) terms containing a bilinear in the antiquark fields as well. The third kind (iii)
corresponds to operators bilinear in both heavy quark and
heavy antiquark fields (four fermion terms).

\medskip

A crucial step in building an EFT for heavy quarks is the so called matching.
In the process of matching we enforce the effective theory to reproduce
suitable S-matrix elements of the full theory. In this way we fix the
parameters (Wilson coefficients) of the effective theory.
Through the matching process the high energy contributions are
encoded in Wilson coefficients multiplying the operators in the Lagrangian (and in the
currents) of
the effective
theory.
The determination of some of these Wilson coefficients of the NRQCD
Lagrangian is the main topic of this paper.

\medskip

The question arises whether
 the Wilson coefficients of the
terms (i) and (ii) in HQET and NRQCD are the same. We shall support below the claim in
\cite{Manohar} that this is
indeed the case. Therefore, since the mass of the heavy quarks is (by
 definition) much larger than $\Lambda_{QCD}$, the matching may be
 done order by order in $1/m$ and $\al_s$.

\medskip

The matching for NRQCD has been known at tree level since
long. This can be obtained by enforcing the tree level of S-matrix
elements to be equal to those of QCD (QED) as mentioned above. For
terms bilinear in the quark (antiquark) fields, this is equivalent to
performing a Foldy-Wouthuysen transformation in the QCD Lagrangian.
Although for HQET the matching at tree level for the bilinear terms can be carried out exactly as above
\cite{Pirjol},
in most of the works
it has been done somewhat differently:
either by imposing the off-shell Green functions be equal to those of
QCD \cite{R1} (see also \cite{Tzani})
or by
integrating out the 'antiparticle' degrees of freedom \cite{Mannel1}.
The Lagrangian obtained in this way is
in fact
 different from the
 NRQCD Lagrangian. However both Lagrangians are related
by local field
redefinitions or by using the equations of motion \cite{Manohar}.
Results for the matching at one loop
 have also been known in the
 HQET for some time \cite{Neubert}.
Nevertheless, attempts to perform the matching
beyond tree level in NRQCD have not begun until recently. The main obstacle
was that in NRQCD, unlike in HQET, the kinetic term was thought to be a necessary
ingredient
in
the quark propagator for a matching calculation,
\be
{1 \over k_0+i \e} \longrightarrow {1 \over \displaystyle k_0 - {
\strut {\bf k}^2
\over \displaystyle 2m}
+i
\e}
\,.
\ee
If a hard cut-off is used ($\mu
<< m$),
it can easily be seen that the matching can be performed just like in
HQET since $k^0>> {\bf k}^2/m$ in the ultraviolet.
However,
if dimensional regularization
is used,
the high energy modes ($k > m$) are not explicitly suppressed and they
give non-vanishing contributions.
This
can be seen because the behavior of the NRQCD propagator changes at
energies larger than the mass. In spite  of this, one would like to use
dimensional regularization because it keeps all the symmetries of QCD and, moreover, the calculations are
technically simpler.
 Several authors have addressed this problem \cite{otros} and
 recently an appealing solution has been proposed \cite{Manohar}. There, it
is claimed that the matching in NRQCD using dimensional regularization should be performed just like in HQET,
namely
the kinetic term must be treated as a perturbation. Let
us
make some remarks which support this approach. The key point is that in order to carry
out the matching
 it is not so important to know the power counting of each term in the effective
theory as
to know that the dynamical scales of the effective theory are much lower than
the
mass. The power counting tells us the relative importance between
different operators but this does not change the value of the matching
coefficients. That is, we only need

\be
m >> |{\bf p}|, E, \Lambda_{QCD}
\ee
no matter what the relation between $|{\bf p|}$, $E$ and $\Lambda_{QCD}$ is.
The above becomes clear if one thinks of the matching as a procedure to integrate out
high energy degrees of freedom {\`a} la Wilson: the effective Lagrangian that we obtain
after integrating
out
energies
and momenta until a scale $\mu$, $ m >> \mu >> |{\bf p}|, E, \Lambda_{QCD}$ does not
depend on the relative weight of the lower scales.

\medskip

In addition, in ref. \cite{Manohar} dimensional regularization was used
to regulate  both the ultraviolet (UV) and the infrared (IR) divergences in the full and
the
effective theory \cite{EH}. The latter arise when the S-matrix elements are expanded about the
residual momentum. In fact, it is not so important to know the way the UV
divergences of the full theory are regulated since the comparison is
done between S-matrix elements which are UV finite (after renormalization).
 Nevertheless, it is
essential to regulate in the same way the IR divergences in both
the full and effective theory in order for them to cancel out. This will always
 happen since by
construction
both theories have
the same IR behavior. It is also important, from a practical
point of view, to regulate the UV
divergences of the effective theory using dimensional regularization.
In this way,
the calculation in the effective theory becomes
trivial since there is no dimensionfull parameter in any integral.
In the ref. \cite{Manohar},  the matching was performed at one loop until $O(1/m^2)$
for operators bilinear in the quark fields.
 It is the aim of this paper to perform the matching at one loop until $O(1/m^2)$
for four-quark operators
and hence
to complete the matching at one loop $O(1/m^2)$.

\medskip

We are thus faced with the computation of S-matrix elements of
four heavy quarks in QCD and HQET.
The computation of these matrix elements in HQET is unusual,
although some related calculations already exists in the literature
\cite{Mannel,nos2}.
Indeed, for heavy-light systems four fermion operators are relevant only when two of the
fermions are light. For heavy quarkonium systems instead all four
quark fields are heavy.
In fact it is in these S-matrix elements where we can
see
the peculiar
IR behavior of heavy-heavy systems, which eventually gives rise to the
 Coulomb pole and hence to the standard non-relativistic weak coupling
bound states.

\medskip

This IR behavior should appear in both the full and the effective theory.
If we
expand about the residual momentum the matrix
elements of the dimensionally regulated QCD, we may expect an IR singularity
reflecting the Coulomb pole. However this singularity corresponds to an odd
 power-like IR
divergence and hence it is put to zero in dimensional regularization. This is not a
problem. Indeed, since the effective theory has the same IR
behavior, it also has an IR divergence reflecting the Coulomb pole which is
consistently put to zero by dimensional regularization.
The important thing when doing the matching is to take into account all the
non-analytical behavior in the heavy quark masses which can not be obtained in the
effective theory. Proceeding
in this way we
are certainly taking into account all the non-analytical behavior in the masses coming
from high momenta (QCD logs).
The remaining non-analytical behavior (Coulomb pole) is encoded in the effective theory.

\medskip

 Although we have been talking about QCD and NRQCD,
the results for
QED and NRQED follow  trivially from our calculations.

\medskip

Let us finally mention some of the possible applications of this work.
The unequal mass case may be important for the $B_c$ system (this
system has been studied in refs. \cite{Sanchis})
which is expected to be seen in the future. This case is
also important in QED for the muonium or Hydrogen-like atoms.
For the equal mass case, our results fixes the scale of the $\als$ running
constant for annihilation contributions to the four quark interaction.
 This is important since in QCD,
at the scales of Bottomonium
and Charmonium, $\als$ strongly depends on the scale. Moreover, since
 there are many scales in the game ($m, {\bf p}, E$) it is not a priori clear which
one should be used in order to fix the value
of $\als$ in the perturbative \cite{yndnos6,PY} and non-perturbative potentials \cite{potential}.
In fact, depending on where the contribution comes from, this value may be
different.
Recently, the spectrum of $\Upsilon (1s)$ and $J/\psi$ has been obtained from
perturbative QCD at $O(m\als^4)$ \cite{PY}. Next improvement, namely $O(m\als^5)$
receives contributions from the matching of four quark operators, and hence our
calculation becomes relevant. It should also be taken into account in
parameterizations of the non-perturbative heavy quark potential along the lines
of reference \cite{potential}.

\medskip

We distribute
the paper as follows. In sec. 2 we define our four quark operators and their Wilson
coefficients. In sec. 3 we
 calculate the Wilson coefficients for the unequal mass case.
In sec. 4 we calculate the Wilson coefficients for equal mass case.
In sec. 5 we discuss a few relevant issues in our calculation. The last
section is
devoted to the conclusions. A few technical points concerning the Coulomb singularity
are
relegated to an Appendix.

\bigskip

\section{Setting the matching}
\indent

\medskip

The piece of the NRQCD
Lagrangian containing four quark operators at $O(1/m^2)$ reads
\bea
\label{lag1}
\delta {\cal L}_{NRQCD} &=&
  {d_{ss} \over m_1 m_2} \psi_1^{\dag} \psi_1 \chi_2^{\dag} \chi_2
+
  {d_{sv} \over m_1 m_2} \psi_1^{\dag} {\bfsigma} \psi_1
                         \chi_2^{\dag} {\bfsigma} \chi_2
\nonumber
\\
&&+
  {d_{vs} \over m_1 m_2} \psi_1^{\dag} {\rm T}^a \psi_1
                         \chi_2^{\dag} {\rm T}^a \chi_2
+
  {d_{vv} \over m_1 m_2} \psi_1^{\dag} {\rm T}^a {\bfsigma} \psi_1
                         \chi_2^{\dag} {\rm T}^a {\bfsigma} \chi_2
\,,
\eea
where $\psi$ is the Pauli spinor field that annihilates a heavy
quark and $\chi$ is the Pauli spinor field that creates a heavy
anti-quark. The subindices 1,2 denotes the possibility of working with
different particles (different masses). We will omit these indices when
the particle-antiparticle case is treated.

\medskip

There is another possibility of writing down these terms by using
Fiertz transformations. It reads

\bea
\label{lag2}
\delta {\cal L}_{NRQCD} &=&
  {d_{ss}^c \over m_1 m_2} \psi_1^{\dag} \chi_2 \chi_2^{\dag} \psi_1
+
  {d_{sv}^c \over m_1 m_2} \psi_1^{\dag} {\bfsigma} \chi_2
                           \chi_2^{\dag} {\bfsigma} \psi_1
\nonumber
\\
&&+
  {d_{vs}^c \over m_1 m_2} \psi_1^{\dag} {\rm T}^a \chi_2
                         \chi_2^{\dag} {\rm T}^a \psi_1
+
  {d_{vv}^c \over m_1 m_2} \psi_1^{\dag} {\rm T}^a {\bfsigma} \chi_2
                         \chi_2^{\dag} {\rm T}^a {\bfsigma} \psi_1
\,.
\eea

The relation between the two bases is

\bea
\nonumber
d_{ss} &=& -{d_{ss}^c \over 2N_c}- {3 d_{sv}^c \over 2N_c}
      -{N^2_c-1 \over 4N_c^2}d_{vs}^c - 3{N^2_c-1\over 4N_c^2} d_{vv}^c
\,,\\
\nonumber
d_{sv} &=& -{d_{ss}^c \over 2N_c}+ { d_{sv}^c \over 2N_c}
      -{N^2_c-1 \over 4N_c^2}d_{vs}^c + {N^2_c-1\over 4N_c^2} d_{vv}^c
\,,\\
\nonumber
d_{vs} &=& -d_{ss}^c - 3 d_{sv}^c
      +{d_{vs}^c \over 2N_c}+  {3 d_{vv}^c \over 2N_c}
\,,\\
d_{vv} &=&  -d_{ss}^c +  d_{sv}^c
      +{d_{vs}^c \over 2N_c}- { d_{vv}^c \over 2N_c}
\,.\eea

Of course, one can always use a redundant bases with the eight operators (2.1) and
(2.2).
The Lagrangian (2.2) is more convenient, as far as the matching calculation is
concerned, when one is dealing with
annihilation
processes in the equal mass case.
Nevertheless
(2.1) is a better option when one addresses a bound state calculation. We shall
use (2.1) for the unequal mass case and the redundant basis for the equal mass one in
order to ease comparison with the actual calculations.

\medskip

In the QED case we have
\be
\label{lag1qed}
\delta {\cal L}_{NRQED} =
  {d_{s} \over m_1 m_2} \psi_1^{\dag} \psi_1 \chi_2^{\dag} \chi_2
+
  {d_{v} \over m_1 m_2} \psi_1^{\dag} {\bfsigma} \psi_1
                         \chi_2^{\dag} {\bfsigma} \chi_2
\,,
\ee
\be
\label{lag2qed}
\delta {\cal L}_{NRQED} =
  {d_{s}^c \over m_1 m_2} \psi_1^{\dag} \chi_2 \chi_2^{\dag} \psi_1
+
  {d_{v}^c \over m_1 m_2} \psi_1^{\dag} {\bfsigma} \chi_2
                           \chi_2^{\dag} {\bfsigma} \psi_1
\,.
\ee
Now, the relation between the two bases is
\bea
\nonumber
d_{s} &=& -{d_{s}^c \over 2}- {3 d_{v}^c \over 2}
\,,\\
d_{v} &=& -{d_{s}^c \over 2}+ { d_{v}^c \over 2}
\,.\eea

 We shall expand the
dimensionally regulated matrix elements about zero residual momentum. Since there are
no derivative terms in (2.1) and (2.2), the zeroth order in the expansion will be
enough. Namely we only have to calculate the matrix element for the four quarks at
rest. This means that the amputated legs in a digram only have to be multiplied
either by
$p_{+}$ (projector on the particle subspace) or $p_{-}$ (projector on the
antiparticle subspace), and the kinematic factor $\sqrt{m/E}$ relating relativistic and
non-relativistic normalizations can be put to one.
  We shall use $\overline{ MS}$ subtraction scheme for both UV and IR divergences
and work with the Feynman gauge. The matching coefficients should be gauge
independent, but they depend on the subtraction scheme. It is worth emphasizing
that we do not work in the on-shell renormalization scheme for the wave function (of
course our masses always correspond to the pole mass), but just $\overline{ MS}$. In
 this scheme (also in $MS$ or similars) the matching can be carried out
straightforwardly.
 If the on-shell scheme is
used for the full or effective theory, one must identify the UV divergences which
correspond to a wave function renormalization and subtract them accordingly (not just
minimally). This is obviously more tedious than using $\overline{MS}$ throughout. The
little price to be paid for this simplicity is that our fields are not properly
normalized. This must be taken into account by including the proper $Z$ factors when
calculating on-shell matrix elements.
\be
Z^{QCD}=  1 + C_f {\als \over \pi} \left( {3 \over 4}\ln{m^2\over \nu^2} -1
\right) + O\left(\left({\als \over \pi}\right)^2 \right)
\,,\quad \quad Z^{NRQCD}=1 \,,\quad \quad C_f = {N^2_c-1 \over 2N_c} \,.
\ee

Notice finally that at the order we are working at the Wilson coefficients in (2.1)
 and
(2.2) are invariant under the local field redefinitions discussed in \cite{Manohar}.

\section{Unequal mass case}
\indent
\medskip

\medskip
\begin{figure}
\hspace{-.1in}
\epsfxsize=3.6in
\centerline{\epsffile{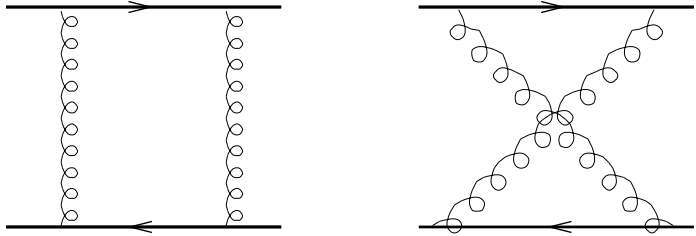}}
\caption {We show the relevant diagrams for the matching of the four-fermion
operators at order $O(1/m^2)$ and one loop for the unequal mass
case. The incoming and outcoming particles are on-shell and exactly at
rest.}
\label{figunmass}
\end{figure}

In this case annihilation diagrams are forbidden and we are only left with the two
QCD diagrams of fig. 1.
In these diagrams
the Coulomb singularity
can be identified and the above mentioned mechanism by which it disappears uncovered.
We show this in detail in the Appendix. In short, things go as follow.
In order to perform some integrals we have to
move to dimensions high enough in order to regulate the IR Coulomb
singularity. When coming back to four dimensions we can trace back the
IR
Coulomb singularity as a pole in higher dimensions,  which does not appear
in four dimensions since dimensional regularization loses power-like divergences.
 The point is that we have not provided a suitable dimensionfull
parameter (the relative momentum) and hence dimensional regularization has no
way to reproduce the Coulomb pole. This fact was pointed out some time ago in
 ref.
\cite{nos2}.

\medskip

We obtain the following matching coefficients

\be
d_{ss}=
  - C_f \left({C_A \over 2} -C_f \right)
    {\al_s^2 \over m_1^2-m^2_2}
\left\{m_1^2\left(  \ln{m^2_2 \over  \nu^2}
                   + {1 \over 3} \right)
       -
       m^2_2\left(  \ln{m^2_1 \over  \nu^2}
                   + {1 \over 3} \right)
\right\}
\,,\ee
\be
d_{sv}= C_f \left({C_A \over 2} -C_f \right)
   {\al_s^2 \over m_1^2-m^2_2}
m_1 m_2\ln{m^2_1 \over m^2_2}
\,,\ee
\bea
\label{dvs}
d_{vs}&=&
- {2 C_f \als^2 \over m_1^2-m^2_2}
  \left\{m_1^2\left( \ln{m^2_2 \over \nu^2}
                   + {1 \over 3} \right)
       -
       m^2_2\left(  \ln{m^2_1 \over  \nu^2}
                   + {1 \over 3} \right)
  \right\}
\\
\nonumber
&&+ { C_A \als^2 \over 4 (m_1^2-m^2_2)}
 \Biggl[
  3\left\{m_1^2\left( \ln{m^2_2 \over \nu^2}
                   + {1 \over 3} \right)
       -
       m^2_2\left(  \ln{m^2_1 \over \nu^2}
                   + {1 \over 3} \right)
  \right\}
\\
&&
\nonumber
\quad\quad\quad\quad
  +
   { 1 \over m_1m_2}
   \left\{m_1^4\left( \ln{m^2_2 \over \nu^2}
                   + {10 \over 3} \right)
       -
       m^4_2\left(  \ln{m^2_1 \over  \nu^2}
                   + {10 \over 3} \right)
  \right\}
 \Biggr]
\,,\eea
\bea
\label{dvv}
&&d_{vv}=
 {2 C_f \als^2 \over m_1^2-m^2_2}
        m_1 m_2\ln{m^2_1 \over m^2_2}
\\
\nonumber
&&+
{ C_A \als^2 \over 4 (m_1^2-m^2_2)}
    \Biggl[
  \left\{m_1^2\left( \ln{m^2_2 \over  \nu^2}
                   + 3 \right)
       -
     m^2_2\left(  \ln{m^2_1 \over \nu^2}
                   + 3 \right)
  \right\}
    -
  3 m_1 m_2\ln{m^2_1 \over m^2_2}
    \Biggr]
\,,
\eea
where $C_A=N_c$.

\medskip

In the case of QED our results reduce to
\be
d_{s}=
  - {\al^2 \over m_1^2-m^2_2}
\left\{m_1^2\left(  \ln{m^2_2 \over  \nu^2}
                   + {1 \over 3} \right)
       -
       m^2_2\left(  \ln{m^2_1 \over  \nu^2}
                   + {1 \over 3} \right)
\right\}
\,,\ee
\be
d_{v}=
   {\al^2 \over m_1^2-m^2_2}
m_1 m_2\ln{m^2_1 \over m^2_2}
\,.\ee

The spin dependent piece, which is subtraction point independent,
  agrees with the result
obtained by Caswell and Lepage in ref. \cite{Lepage}. The scalar piece
is new.

\medskip

Since we will need the equal mass results in the next section let us
display them here. For QCD they read

\be
d_{ss}=
  -
C_{f}\left( {C_{A}\over 2}-C_{f}\right)
 \al_s^2
    \left(  \ln{m^2 \over  \nu^2}
                   - {2 \over 3} \right)
\,,\ee
\be
d_{sv}=
C_{f}\left( {C_{A}\over 2}-C_{f}\right)
 \al_s^2
\,,\ee
\be
d_{vs}=
  - 2 C_f \al_s^2
    \left(  \ln{m^2 \over  \nu^2}
                   - {2 \over 3} \right)
  + {5 \over 4} C_A \al_s^2
    \left(  \ln{m^2 \over  \nu^2}
                   + {11 \over 15} \right)
\,,\ee
\be
d_{vv}=
   2 C_f \al_s^2
  + {  C_A \al_s^2\over 4}
    \left(  \ln{m^2 \over  \nu^2}
                   -1 \right)
\,.\ee

For QED we have
\be
d_{s}=
  - \al^2
    \left(  \ln{m^2 \over \nu^2}
                   - {2 \over 3} \right)
\,,\ee
\be
d_{v}=
   \al^2
\,.\ee

Recently, the scalar piece for the equal
mass case in QED was calculated in ref. \cite{Labelle2} using a
cut-off regularization. This result agrees with ours except for a finite piece, which
may be
due to a different renormalization scheme for the four fermion operators in the
effective theory.
This contribution is relevant for
the full calculation of the positronium energy levels at order
$O(m\alpha^5)$. Work in this direction is under way \cite{nosn}.

\bigskip

\section{Equal mass case}
\indent
\medskip

\medskip
\begin{figure}
\hspace{-.1in}
\epsfxsize=4.2in
\centerline{\epsffile{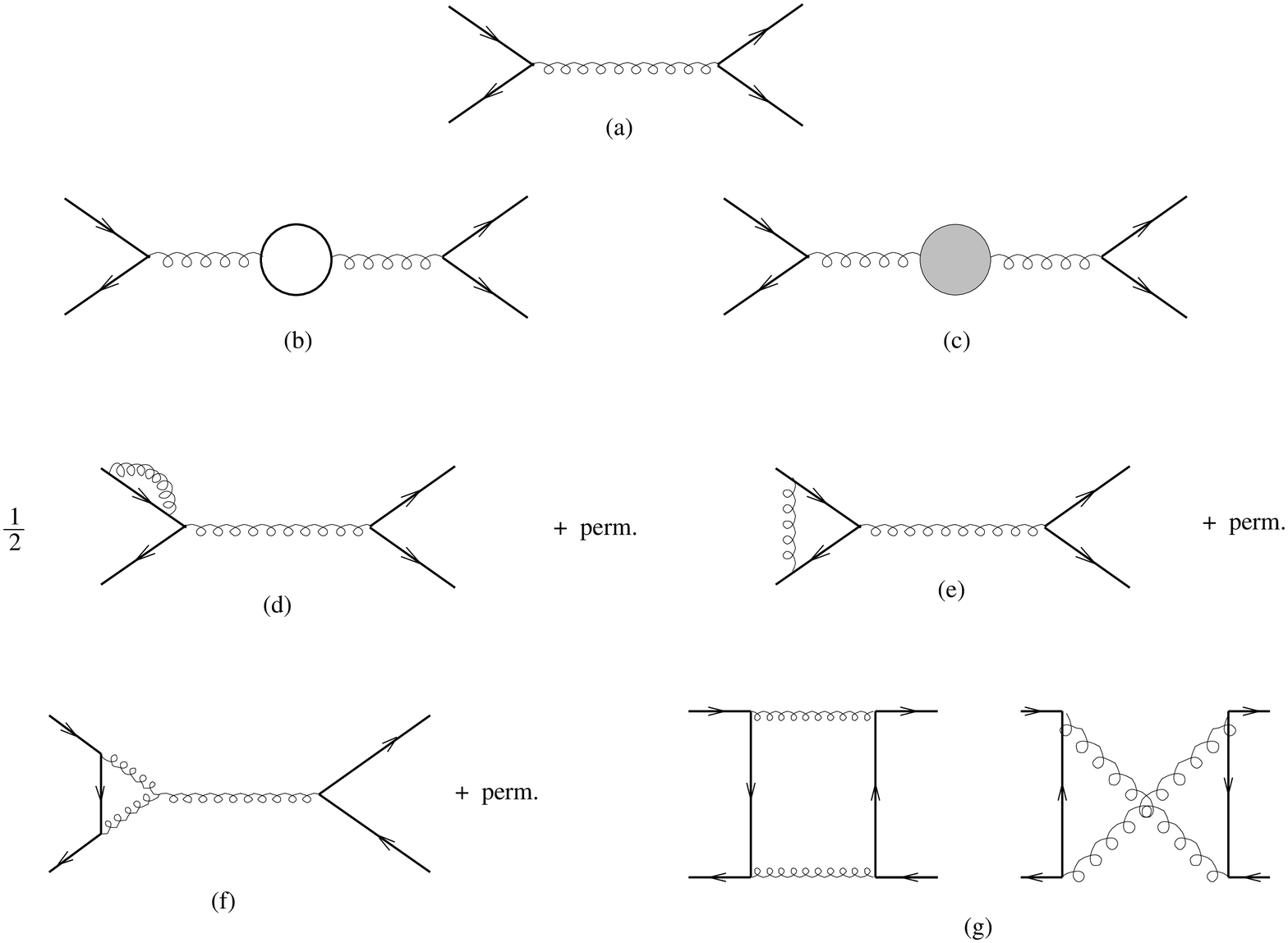}}
\caption {We show the relevant diagrams to the matching for the four-fermion
operators at order $O(1/m^2)$ and one loop that only appear for the
equal mass case. The incoming and outcoming
particles are on-shell and exactly at rest.}
\label{figeqtree}
\end{figure}

For equal particles annihilation processes are allowed and they should
be taken into account (fig. 2).
From fig. 2a we obtain the well known result \cite{Lepage2}

\be
d_{vv}^{c,2a}= -\pi \als
\,.\ee

This is the lower order (tree level) contribution.
Let us consider first the one loop contributions arising from the gluon self-energy.
Each heavy
quark loop (fig. 2b) gives a contribution
\be
d_{vv}^{c,2b}=(-\pi \als)T_R \left(- {8 \als \over 9 \pi}
+{ \als  \over 3 \pi}
\ln{m^2 \over \nu^2}
\right)
\,,\ee
where $T_R=1/2$ for QCD and $T_R=1$ for QED. The QED result had been
already obtained in refs. \cite{Labelle2,Kinoshita}. Light quarks ($n_f$) and
gluons give a contribution (fig. 2c)
\be
d_{vv}^{c,2c}=(-\pi \als) \left(-{\als \over \pi} \right)
\left\{ {C_A \over 4}
  \left( {5 \over 3} \ln{-4m^2-i\e\over \nu^2} -{31 \over 9} \right)
         -
        {T_R \over 3}n_f
  \left( \ln{-4m^2-i\e\over \nu^2} -{5 \over 3} \right)
\right\}
\,.
\ee

\medskip

The quark self-energy diagrams (fig. 2d) do contribute to the matching at
$O(1/m^2)$ in the $\overline{MS}$ scheme, even though they do not in the on-shell
scheme.
The matching coefficient reads
\be
d_{vv}^{c,2d}=4\times (-\pi \als)
{ C_f\over 2} {\als \over \pi} \left( {3 \over 4}\ln{m^2\over \nu^2} -1 \right)
\,.\ee

Let us next consider the vertex corrections (figs. 2e and 2f).
Fig. 2e is quite interesting.
A singularity associated with
the Coulomb
pole should appear, but again it does not show up when doing the computation in
dimensional
regularization for quarks at rest. This is totally analogous to what happened with
diagrams in fig. 1 in the previous section. No signal of infinity
imaginary anomalous dimension appears either \cite{Mannel}.
We refer the reader to ref. \cite{nos2} for a detailed explanation on what is going on.
 We obtain
\be
d_{vv}^{c,2e}=2\times(-\pi \als)
\left({C_{A}\over 2}-C_{f}\right) {3 \als \over 4 \pi}
  \left( \ln{m^2\over \nu^2} +{4 \over 3} \right)
\,.
\ee

For QED fig. 2e has been already computed \cite{Labelle2, Kinoshita}.

\medskip

The diagrams of fig. 2f do not appear in QED.
They lead to
\be
d_{vv}^{c,2f}=2\times (-\pi \als) \left(-{3 \als \over 4 \pi} \right)
{C_A \over 2}
  \left( \ln{m^2\over \nu^2} -{8 \over 9}\ln2 -{16 \over 9} +
{4 \over 9} i \pi
   \right)
\,.
\ee

\medskip

 Finally we consider the contributions from the diagrams of fig. 2g. These diagrams
also exist in QED and their contributions in this theory
have already been calculated in ref.
\cite{Labelle2}.
We obtain from the diagrams of fig. 2g
\be
d_{ss}^{c,2g}= \als^2
C_{f}\left( {C_{A}\over 2}-C_{f}\right)
              \left(2-2\ln2 + i \pi\right)
\,,\ee
\be
d_{vs}^{c,2g}= {\als^2 \over 2}
               \left(-{3 \over 2}C_A+4C_f \right)
                \left(2-2\ln2 + i \pi\right)
\,,\ee
\be
d_{vv}^{c,2g}= \als^2
               {C_A \over 2}
      \left\{ \ln{m^2\over \nu^2} +
               {1 \over 6}
                \left(2-2\ln2 + i \pi\right)
      \right\}
\ee
and $d_{sv}^{c,2g}$ is zero. For QED we reproduce the results in
\cite{Labelle2,Kinoshita}.

\medskip


Summarizing all the contributions from annihilation diagrams
we obtain
\bea
d_{ss}^{c}&=& \als^2 C_{f}\left({C_{A}\over 2}-C_{f}\right)
               \left(2-2\ln2 + i \pi\right) \,,\\
d_{sv}^{c}&=& 0 \,,\\
d_{vs}^{c}&=& {\als^2 \over 2}
               \left(-{3 \over 2}C_A+4C_f \right)
                \left(2-2\ln2 + i \pi\right) \,,\\
d_{vv}^{c}&=& (-\pi\al_{s})\Biggl[1+{\als\over\pi}\Biggl(
T_{R} \left[{1\over 3}n_{f}\left( \ln{m^2\over \nu^2}+2\ln2-{5\over 3}-i\pi\right)
-{8\over
9}+{1\over 3}\ln{m^2\over \nu^2}\right] \nonumber\\
&  &+C_{A}\left[ -{11\over 12}\ln{m^2\over \nu^2} +{109\over 36}
\right] + C_{f}\left[ -4\right]\Biggr)\Biggr]
\,.\eea
Recall that we have to add to the annihilation contributions above the contributions
(3.7)-(3.10).

\medskip

For QED we have
\bea
d_{s}^{c}&=& \al^2
               \left(2-2\ln2 + i \pi\right) \,,\\
d_{v}^{c}&=& (-\pi\al)\Biggl[1+{\al\over\pi}\Biggl(
-{44\over
9}+{1\over 3}\ln{m^2\over \nu^2}
\Biggr)\Biggr]
\,.\eea

\bigskip
\section{Discussion}

Let us first address the important question on how the matching calculation helps to
fix the scale of $\als$. In the previous section we have not paid any attention to the
flavor dependence of $\als$. For simplicity let us focus on the case of a
single heavy flavor. Suppose that in QCD we have $N_{f}$ flavors. Then in NRQCD we
have $n_f=N_{f}-1$ relativistic flavors. Consequently, the $\overline{MS}$
 running coupling
constant in NRQCD is expected to run according to $N_{f}-1$ flavors.
However, this is not obvious from the matching calculation.
Notice that the $\als$s in the
NRQCD Lagrangian (both explicit and in the Wilson
coefficients) are those inherited from QCD and hence one may be tempted to make them
run with $N_{f}$ flavors. In order to clarify this issue consider first the pure
gluonic part of the NRQCD Lagrangian \cite{Manohar}.
\be
\label{laggl}
{\cal L}=
-{1 \over 4}d_1 G^A_{\mu\nu} G^{A\mu\nu}
+{d_2 \over m^2} G^A_{\mu\nu} D^2 G^{A\mu\nu}
+{d_3 \over m^2} g f_{ABC}G^A_{\mu\nu} G^B_{\mu\al} G^C_{\nu\al}
\,,
\ee
where
\begin{eqnarray}
d_1 &=& 1-{ \als\over 3 \pi} T_{R} \ln m^2/\nu^2, \nonumber \\
d_2 &=& {\als\over 60 \pi} T_{R},\\
d_3 &=& {13\als\over 360 \pi} T_{R}\,.
\nonumber
\end{eqnarray}
Notice that the kinetic term does not have the standard normalization anymore. This can
be recovered by a simple redefinition of the gluon field. Since the remaining gluon
fields in the NRQCD Lagrangian are multiplied by $g$, this is equivalent to make the
change
\be
g\rightarrow g\left( 1-{ \als\over 3 \pi} T_{R} \ln m^2/\nu^2
\right)^{-{1\over 2}} \Longleftrightarrow \als \rightarrow \als\left( 1-{ \als\over 3
 \pi} T_{R} \ln m^2/\nu^2
\right)^{-1}
\ee
in all the $g$s which multiply the gluon fields. At one loop this is nothing but
changing the running coupling constant of $N_{f}$ flavors for the running coupling
constant of $N_{f}-1$ flavors which is a desired result. However there are
additional dependences on $\als$ in the NRQCD Lagrangian (which are not multiplying
gluon fields) in the Wilson coefficients.
 Notice however that the
difference between $\als^{N_{f}}$ and $\als^{N_{f}-1}$ is higher order in $\als$ and
hence we can safely substitute $\als^{N_{f}}$ by $\als^{N_{f}-1}$ in all Wilson
coefficients except in (4.13). Indeed, the Wilson coefficient in (4.13) is linear
 in $\als$ and hence
it may be sensitive on whether this $\als$ corresponds to $N_{f}$ or $N_{f}-1$ flavors. Since
this $\als$ is inherited from QCD it corresponds to $N_{f}$ flavors.
However the $\n$ dependence enters in such a way that
\bea
\nonumber
&&\al_{s}^{N_{f}}(\n)\Biggl[1+{\als^{N_{f}} (\n) \over\pi}\Biggl(
T_{R} \left[{1\over 3}n_{f} \ln{m^2\over \nu^2}
+{1\over 3}\ln{m^2\over \nu^2}\right]
+C_{A}\left[ -{11\over 12}\ln{m^2\over \nu^2}
\right] \Biggr)\Biggr]
\\
&&
=\als^{N_{f}} (m)=\als^{N_{f}-1} (m)
\eea
and hence the scale of $\als$ is naturally fixed to $m$. The Wilson coefficient $d_{vv}^{c}$
in (4.13) should better be written like
\bea
d_{vv}^{c}&=& (-\pi\al_{s}(m))\Biggl[1+{\als\over\pi}\Biggl(
T_{R} \left[{1\over 3}n_{f}\left( 2\ln2-{5\over 3}-i\pi\right)
-{8\over
9}\right]\nonumber \\
&  &+C_{A}\left[ {109\over 36}
\right] + C_{f}\left[ -4\right]\Biggr)\Biggr]
\,.\eea
Therefore, we have seen that
the $\als$ in the NQRCD Lagrangian correspond
to running coupling constants at two different scales. The $\als$ multiplying the gluon
fields must be understood at some scale $\n$, $\n << m$ and run according to $N_{f}-1$
relativistic flavors, whereas the $\als$ in the Wilson
coefficients must be understood at the scale $m$.

\medskip

Let us next comment on the case of two different heavy flavors. If one takes the
Wilson renormalization group point of view strictly,
matching QCD with $N_{f}$ flavors to NRQCD with $N_{f}-2$ relativistic flavors
makes sense only if $m_1\sim m_2$. If, say, $m_1 >> m_2$, one should better do the
matching in two steps. First one should match QCD to NRQCD with $N_{f}-1$ relativistic
flavors (NRQCD$_{N_{f}-1}$) and next NRQCD$_{N_{f}-1}$ to NRQCD$_{N_{f}-2}$.
Nevertheless, if there is no dynamical scale between $m_1$ and $m_2$ and we are not
interested in any renormalization group improvement of the Wilson coefficients,
carrying out the matching in one step or in two steps must give exactly the same
result to any
fixed order in perturbation theory. Then, we expect our results to be useful for the
$B_c$ meson in QCD and for the muonium and hydrogen-like atoms in QED.
Recall that the Wilson coefficients in (5.2) trivially change into
\begin{eqnarray}
d_1\longrightarrow & d_1 = 1-{ \als\over 3 \pi} T_{R} \left(\ln m_1^2/\nu^2+\ln
m_2^2/\nu^2\right)
 \nonumber \,,\\
{d_2\over m^2}\longrightarrow &\displaystyle{ {d_2\over m_1^2}+{d_2\over m_2^2}}
\,,\\
{d_3\over m^2}\longrightarrow &\displaystyle{ {d_3\over
m_1^2}+{d_2\over m_2^2}} \,.
\nonumber
\end{eqnarray}
Now rescaling the gluon field to its usual normalization ($d_1=1$) is equivalent to
\be
 \als \rightarrow \als\left( 1-{ \als\over 3
 \pi} T_{R} \left(\ln m_1^2/\nu^2+\ln m_2^2/\nu^2\right)
\right)^{-1}
\ee
in the coupling constants multiplying the gluon fields, and hence these $\als$ run
with $N_{f}-2$ flavors. Notice also that the following equality holds,
$\als^{N_{f}} (\nu=\sqrt{m_1 m_2})=\als^{N_{f}-2} (\nu=\sqrt{m_1 m_2})$.

\medskip

For QED an analogous discussion implies that (4.15) is given in terms
of the QED running
coupling constant. This expression in terms of the low energy $\alpha$ ($\alpha\sim
1/137$) reads
\be
\label{5.8}
d_{v}^{c}= (-\pi\al)\Biggl[1+{\al\over\pi}\Biggl(
-{44\over
9}
\Biggr)\Biggr]
\,.\ee
If we add (3.11) and (3.12) to (4.14) and (\ref{5.8}) respectively, we
obtain the results presented in \cite{Mont}.

\medskip

Imaginary parts appear in the Wilson coefficients at several instances. In order to
obtain them from our expressions beware that we have
located the cut at the negative real axes of the $m^2$ complex plain. These imaginary
parts have to do with inelastic cross sections which cannot be obtained within the
non-relativistic theory alone. They are also related to the decay width of
heavy quarkonium states into light hadrons and had been calculated before
\cite{Lepage}. Our results agree with this previous calculation.

\medskip

A word of caution is required when dealing with the Pauli matrices in
$D$ dimensions. 
The Pauli matrices arising in NRQCD have in fact very
different origin, as we comment next. For the non-annihilation diagrams the Pauli
matrices originate from
\be
p_{+}\s^{\m\n} p_{+}\otimes p_{-}\s_{\m\n} p_{-} =: -{1 \over 4}
[\s^i,\s^j] \otimes [\s^i,\s^j] = (D-2)\s^{k}\otimes\s^{k}
\,.\ee
While the first equality can be understood as a definition, for the
second one we have used the following prescriptions (with the proper
limit when $D \rightarrow 4$)
\be
[\s^ i,\s^ j]=2i \e^{ijk} \s^k \,,\quad \quad \e^{ijk}\e^{ijk^{\prime}}
=(D-2)\d^{kk^{\prime}} 
\,.\ee
The finite part of $d_{vv}$ depends on these prescriptions. For the
annihilation diagrams the Pauli matrices originate from\footnote{Notice that $\s^{k}\otimes\s^{k}$ in (5.9)
and (5.11) act on different spaces even though we did not write this
distinction between them explicitly. Recall also that the Fiertz
rearrangements of section 2 only hold in four dimensions as well.} 
\be
p_{+}\g^{\m} p_{-}\otimes p_{-}\g_{\m} p_{+} =: -\s^{k}\otimes\s^{k}
\,.\ee
When carrying out
calculations in dimensionally regulated NRQCD the same prescriptions
have to be used and it may be eventually important to keep in mind the different
origin of the various  $\s^{k}\otimes\s^{k}$
and $ 1\otimes 1$. For NRQED there are no ambiguities at
this order since the spin dependent terms are finite.

\medskip

Let us finally mention that the matching at one loop for the electromagnetic current at
leading order in $1/m$ arises trivially from the calculation of the diagrams in figs. 2b
and 2c. Schematically, the full current is
approximated by
\be
J_{em}={\bar \Psi}\gamma^{\mu} \Psi \rightarrow (1+ {\d_{r} \over 2})
\psi^{\dagger}\s^{i}\chi + O({1 \over m^2})
\,,\ee
where $\d_{r}$ encodes the one loop correction due to hard gluons. Now,
one only has to realize that the relevant computation (the matching
procedure follows analogously to the one for the four-fermion operators)
is the one we performed
for the diagrams above
but taking into account the different color factors.
We obtain
\be
\d_{r} = {1\over (-\pi\alpha_s)}\left( d^{c,2d}_{vv} +{ C_f\over  C_f - C_A/2}
 d^{c,2e}_{vv}\right)= - 4 C_f {\als \over \pi}\,,
\ee
which agrees with the well known result.
We consider this procedure by far the simplest and most efficient method
to obtain $\d_{r}$ (one can also trivially obtain the result
for QED, $\displaystyle \d_{r} = - 4 \al / \pi$). 
No problem with the Coulomb pole appears
through the calculation.
Notice also that no anomalous dimension appears either. This could be traced back
to the fact that in both QCD and HQET for one quark and one
antiquark (the effective theory to which we are matching to
from a practical point of view) have symmetries which protect this
current. For the effective theory this symmetry is $U(4)$ \cite{TS}.

\bigskip
\section{Conclusions}
\indent
\medskip

We have calculated the matching coefficients of the four quark operators of NRQCD at
one loop and $O(1/m^2)$. We have considered both the unequal and equal mass cases.
We have shown explicitly how some matching coefficients in the NRQCD
Lagrangian conspire in such a way that all $\als$ appearing in them must be considered
at the scale $m$ whereas the $\als$ multiplying the gluon fields must be considered
as running with the number of remaining relativistic flavors only.

\medskip

The binding energies of $\Upsilon (1s)$ and $J/\psi$ have been recently obtained from
perturbative QCD at $O(m\als^4)$ \cite{PY}. Next improvement, namely $O(m\als^5)$
in the NRQCD framework requires the knowledge of the matching coefficient of the
four quark operators at one loop calculated here. In the framework of NRQED, these are
also necessary to obtain the positronium binding energy at  $O(m\al^5)$.

\medskip

The unequal mass case in NRQCD may have eventual applications to the $B_c$ meson. For
NRQED it may be relevant for precision calculations (involving recoil corrections) in
muonium and hydrogen-like atoms. In particular it would be relevant for the spectrum
of an hydrogen atom at $O(m\al^5)$ where the electron has been substituted by a $\tau$
particle.

\medskip

{\bf Acknowledgments}
\medskip

A.P. acknowledges a fellowship from Generalitat de Catalunya.
 Financial support from CICYT, contract AEN95-0590 and financial
support from CIRIT, contract GRQ93-1047 is also acknowledged.

\bigskip


\appendix

\section{Coulomb singularity}

In this appendix we show how the Coulomb singularity is reflected in
our calculation.

\medskip

Consider the following integrals
\be
I_{n}=\int {d^{D}q\over (2\pi)^{D}}{1\over \left(q^2 +i\eta\right)^{n}}{1\over q^2+2mq^0 +i\eta}
{1\over q^2-2mq^0 +i\eta}
\,.\ee
$n=1$ and $n=2$ appear in the calculation of the diagrams in fig. 2e and fig. 1
respectively. Upon integration over $q^0$ we obtain IR singularities from the poles
of the quark propagators and from the poles of the gluon propagators.

\medskip

The poles of the quark propagators produce
the Coulomb singularity
\be
I_{n}^{C}\sim \int_{\Lambda} d^{D-1}{\bf q}\left( {1\over{\bf q^2 }}\right)^{n}
{1\over {\bf q^2}
m}\sim \Lambda^{D-3-2n}
\,,\ee
where $\Lambda\rightarrow 0$ is an IR cut-off. At D=4 this integral has odd power like
singularities which are ignored by dimensional regularization. However we expect these
singularities to shown up as poles in an odd number of dimensions.

\medskip

The poles in the
gluon propagators also give rise to IR singularities. These read
\be
I_{n}^{G}\sim \int_{\Lambda} d^{D-1}{\bf q}\left( {1\over{\bf q^2 }}
\right)^{n-1} {1\over\left( {\bf
q^2}\right)^{3\over 2}
m^2}\sim \Lambda^{D-2-2n}
\,.\ee
For $n=1$ and $n=2$ we expect a pole in $D=4$ and $D=6$ respectively (an extra pole at
$D=4$ cannot be ruled out a priori for $n=2$ but it will not turn up).

\medskip

The explicit result for $I_{n}$ below fulfills the expectations above
\be
I_{n}={i\over (4\pi)^{2}}
\left({-1 \over m^2}\right)^{n}
\left({m^2 \over 4\pi}\right)^{\e}
{\Gamma(n-\e)\Gamma(2\e-2n+1) \over \Gamma(2\e+2-n)}
\ee
where $D=4+2\e$. Notice that the $\e=n$ singularities are of UV origin.

\bigskip



\end{document}